\documentclass[aps,pra,twocolumn,showpacs]{revtex4}

\usepackage{graphicx, amssymb}

\begin{document}

\bibliographystyle{apsrev}

\title{Probability, unitarity, and realism in generally covariant quantum information}

\author{S. Jay Olson}
 \email{solson2@lsu.edu}
 \author{Jonathan P. Dowling}
\affiliation{Hearne Institute for Theoretical Physics, Department of Physics and Astronomy, Louisiana State University, Baton Rouge, LA 70803}

\date{\today}

\begin{abstract}
The formalism of covariant quantum theory, introduced by Reisenberger and Rovelli (RR), casts the description of quantum states and evolution into a framework compatable with the principles of general relativity.  The leap to this covariant formalism, however, outstripped the standard interpretation used to connect quantum theory to experimental predictions, leaving the predictions of the RR theory ambiguous.  Here we discuss in detail some implications of our recently proposed description of covariant quantum information (CQI), which addresses these problems.  We show explicit agreement with standard quantum mechanics in the appropriate limit.  In addition to compatability with general covariance, we show that our framework has other attractive and satisfying features \---- it is fully unitary, realist, and self-contained.  The full unitarity of the formalism in the presence of measurements allows us to invoke time-reversal symmetry to obtain new predictions closely related to the quantum Zeno effect.  
\end{abstract}


\maketitle

\section{Introduction}

Quantum information in recent years has become a topic of enormous interest to the physics community.  Driven primarily by interest in exotic new technologies, the surge in interest has also cast new light on the foundations of quantum mechanics (QM).  There is a similar dual role for relativistic quantum information~\cite{Peres1}.  On the one hand, incorporating aspects of general relativity (GR) into quantum information theory will be needed to extend quantum technology over large regions of spacetime in the presence of gravity \---- for example, between satellites in Earth orbit.  On the other hand, an understanding of generally covariant quantum information may be essential for interpreting quantum gravity, where many standard assumptions taken for granted in quantum mechanics are no longer valid.

In 2002, Reisenberger and Rovelli (RR) described a generally covariant quantum formalism, designed to be applicable to any quantum system, up to and including quantum gravity~\cite{Rovelli1}.  Originally, the main difficulty with this formulation was in correspondence with orthodox, single particle quantum mechanics (QM).  In standard quantum mechanics, the measurement postulate forces the future to be treated very differently from the past; in the covariant approach, this sort of distinction is not allowed.  It should be emphasized that this time asymmetric feature of quantum mechanics does not go away simply by making the dynamics globally or locally Lorentz invariant \---- it has been firmly entrenched within the postulates of quantum mechanics from the beginning.  Additionally, the RR probability postulate relies upon a certain small-region approximation to obtain agreement with standard quantum theory.  As we will see, both of these problems (as well as their solutions, discussed herein) are subtly connected to quantum information.  

The temporal correspondence problem manifests itself most clearly in the case of multiple measurements.  To see this, recall that in standard quantum theory, if we are given a state $|\Psi \rangle$, the probability to observe this system in state $|\phi \rangle$ is given by $| \Pi_{\phi} |\Psi \rangle |^2$, where $\Pi_{\phi}$ is the projector onto the state $|\phi \rangle$.  What is the probability that we observe the system in state $|\phi \rangle$ \emph{and} in state $|\sigma \rangle$?  It is $| \Pi_{\sigma} \Pi_{\phi} |\Psi \rangle |^2$ if we perform the $\phi$-measurement first, and $| \Pi_{\phi} \Pi_{\sigma} |\Psi \rangle |^2$ if we perform the $\sigma$-measurement first.  Since the projectors may not commute, we must specify time ordering (distinguishing past from future) with respect to some causal structure to obtain unique predictions, which can be compared to experiment.  But this kind of causal structure is exactly what covariant quantum theory lacks.  How are we to obtain unique, causally ordered probabilities?

Recently, we have proposed that an information-theoretic approach should be used to obtain predictions from covariant quantum theory~\cite{Olson1} (an approach we refer to as covariant quantum information, or CQI).  Using a generalized partial trace, the CQI postulate immediately reproduces the predictions of standard quantum mechanics (including an effective notion of quantum collapse) in the appropriate single-particle Schr\"odinger-picture limit, but requires no special time variable to be identified to make well-defined multiple-measurement predictions and requires no small region approximation to be invoked in order to obtain agreement with standard QM.  A form of time ordering emerges, nonetheless, as an entropy relationship in the state of the observers, which are modeled quantum mechanically.

In the present paper we explore in more detail the implications of this concept.  In section II, we review the basic elements of RR covariant quantum mechanics, and our information theoretic formalism, CQI.  In section III, we directly compare and contrast the CQI and RR probability postulates.  We analyze a perturbative measurement calculation that demonstrates disagreement between the two approaches, and agreement of CQI with orthodox quantum theory.  In section IV, we discuss a fully realist interpretation of quantum theory allowed by the CQI approach.  In section V, we apply the formalism to the Einstein, Podolsky, Rosen (EPR) scenario and discuss a new framework for understanding nonlocal correlations, without invoking any nonlocal collapse.  In section VI, we take advantage of the unitary framework of this measurement formalism and demonstrate time-reversal in the context of the quantum Zeno effect \---- an effect on the evolution of a quantum system obtained by \emph{disentangling} from an ancilla (in contrast to the standard description of the Zeno effect, which can be understood in terms of projective collapse, or an entangling interaction).  In section VII, we point out the fully self-contained nature of our probability postulate, and discuss the possible implications for quantum cosmology.  We offer our concluding remarks in section VIII.

\section{Covariant Quantum Theory and Information}

Here we review the basic ingredients of our work.  We first consider the Cerf-Adami (CA) measurement formalism of standard, non-relativistic Schr\"odinger-picture quantum mechanics, which is the conceptual starting point of our work~\cite{Adami1, Adami2}.  We then review the RR formalism of covariant quantum theory and the associated problems with correspondence to orthodox quantum theory.  Finally, we show how to merge these ideas into a measurement formalism we call covariant quantum information (CQI), which is free of the correspondence problems.  Section III will be devoted to a direct comparison of CQI with the original RR measurement postulate.

\subsection{The Cerf-Adami Description of Measurement}
Here we begin in the standard, non-covariant Schr\"odinger-picture description of states and evolution.  The states of the theory are generally $L^{2}$ functions of some non-relativistic configuration space $\mathcal M_{0}$.  A unitary time-evolution operator evolves these states from one to another through different instants of the classical background time $t$.

Measurement, however, is not described by the usual non-unitary projection onto an eigenstate of an observable, nor is decoherence by an external environment invoked.  Instead, we model the observer quantum mechanically, describing his interaction with the quantum system unitarily, while keeping track of the entropy of the reduced density operator describing the observer alone \---- this is the central insight of Cerf and Adami~\cite{Adami1, Adami2}.  This approach is motivated by the observation that the outcome of all quantum mechanical experiments can ultimately be phrased in terms of the final state of the observer \---- is he described by a state of $|\mathrm{observed\ outcome}\ i \rangle$, or a state of $|\mathrm{observed\ outcome}\ j \rangle$?  The probabilistic features of quantum mechanics are thus interpreted as components of the mixed state \emph{of the observer}, rather than probabilities for the rest of the universe (except the observer) to make a non-unitary jump.  This subtle shift in focus is extremely important.

To see how this idea works, and reproduces standard physics while retaining unitarity, we consider a sequence of measurements of (in general) incompatible observables \---- an analysis first described in Ref.~\cite{Adami1}.  Consider an arbitrary quantum system we wish to study, $Q$, in the following state:
\begin{eqnarray}
	|Q \rangle = \sum_{i} \alpha_{i} |a_{i} \rangle.
\end{eqnarray}
To perform the first measurement, we allow interaction with a measurement/observer system $A$ (Alice), whose basis states $|i \rangle $ are in one-to-one correspondence with eigenstates $|a_{i}\rangle$ of the observable Alice is measuring, $\hat{a}$.  These states of Alice reflect her realization of a particular outcome of the experiment \---- e.g. the state $|i \rangle$ should be thought of as the state of Alice in which she has observed $Q$ to be in the eigenstate $|a_{i} \rangle$ of the observable $\hat{a}$.  The interaction is such that after it has taken place, we have the following entangled state of $QA$ (Alice was also present before the interaction, but separable with respect to $Q$):
\begin{eqnarray}
	|QA \rangle = \sum_{i} \alpha_{i}|a_{i},i \rangle.
\end{eqnarray}
Now we introduce a second measurement system $B$ (Bob), whose basis states $|j \rangle $ are in one-to-one correspondence with eigenstates $|b_{j} \rangle$ of an observable $\hat{b}$ that does not commute with $\hat{a}$ (implying that the overlap $U_{ij} = \langle b_{j}|a_{i} \rangle$ is not the identity matrix).  After interaction with $B$, we find the system in the following state:
\begin{eqnarray}
	|QAB \rangle = \sum_{i,j} \alpha_{i} U_{ij} |b_{j},i,j \rangle.
\end{eqnarray}
This is a pure state with zero entropy.  However, it contains everything we need to obtain probabilities for the outcome of Alice and Bob's experiment.  Notice that the observers generally do not have access to the full state of the system.  Alice generally only has access to the state of Alice.  Thus, to predict what Alice sees in this experiment, we are interested only in the reduced density operator $\rho_{A}$, and similarly $\rho_{B}$ for Bob.  We do this by performing a partial trace, obtaining:
\begin{eqnarray}
\rho_{A} &=& \sum_{i} |\alpha_{i}|^{2} |i \rangle \langle i|,
	\\
\rho_{B} &=& \sum_{i,j} |\alpha_{i}|^{2} |U_{ij}|^{2} |j \rangle \langle j|.
\end{eqnarray}
Although the full state is pure, the states of the observers themselves are mixed \---- classical probabilities have thus entered the picture.  These density operators have von Neumann entropy identical to the classical Shannon entropy associated with random variables with the following probability distributions:
\begin{eqnarray}
p_{A}(i) &=& |\alpha_{i}|^{2},\\
p_{B}(j) &=& \sum_{i} |\alpha_{i}|^{2} |U_{ij}|^{2}.
\end{eqnarray}
Note that these are exactly the probabilities associated with the traditional assumption of a non-unitary projective collapse during Alice's measurement, and another during Bob's subsequent measurement, though we have introduced no such concept.  If Alice had not performed her initial measurement, Bob's probabilities would have taken the form $p_{B}(j) = | \sum_{i} \alpha_{i} U_{ij}|^{2}$ \---- clearly Alice has induced an \emph{effective} collapse in this description.  This occurs because we are looking only at the state of the observers (hence the partial trace) and the classical probabilities associated with these being described by a particular quantum mixture.  We are free to read off the probabilities directly without bringing non-unitarity into the theory.  

This prescription is quite general.  For a sequence of $N$ measurements, we have the following density operator for the $N^{th}$ observer~\cite{Adami3}:
\begin{eqnarray}
\rho_{N} = \sum_{i,j...k,l} |\alpha_{i}|^{2} |U^{(2)}_{ij}|^{2} ... |U^{(N)}_{kl}|^2 |l \rangle \langle l|.
\end{eqnarray}
Defining the conditional probabilities $p_{n}(ij) = |U^{(n)}_{ij}|^{2}$, this implies the following probability relation:
\begin{eqnarray}
p_{N}(l) = \sum_{i,j...k} p_{1}(i) p_{2}(ij)... p_{N}(kl).
\end{eqnarray}
The form of this probability chain implies an important property, that the entropy of subsequent observers must be at least as great as that of the preceding observers.  This is closely related to a well-known result \---- that projective measurements can only increase or leave constant the entropy of a system~\cite{Nielsen1} [note that $S(Q)$ is always equal to the entropy of the last observer, $S(N)$].  Here, this entropy increase by projective measurements is reflected by the increasing entropy of the observer systems (since the entropy of the \emph{combined} system remains zero).  In this way, a quantum mechanical arrow of time (the direction of increasing observer entropy) as well as an effective quantum mechanical collapse is subtly hidden in Schr\"odinger picture quantum mechanics, when information theory is accounted for \---- we did not have to postulate these things, they are emergent within a fully unitary framework, so long as we remember to interpret probabilistic predictions in terms of the state of the observers, rather than the state of $Q$.

We must mention an objection, sometimes raised against the standard decoherence picture of measurement~\cite{Penrose1}, which leads to a similar mixed density operator method of obtaining probabilities (though the logic and interpretation is quite different \---- here we are concerned with the reduced density operator describing the observer $A$ and specifically \emph{not} the system $Q$ under observation, and we required no external environment to interact with the system or observers).  There are many ways to express the mixed density operator $\rho_{A}$ \---- how do we know which basis corresponds to the \emph{experience basis} of Alice, i.e., what is the basis in which Alice does not find herself in a superposition after the measurement?  Answering this question is essential if we are to obtain the proper probabilities and make unambiguous predictions.

We propose that the preferred representation of $\rho_{A}$, from which probabilities may be read directly, is the \emph{unique} basis in which $\rho_{A}$ is diagonal and whose components are orthogonal.  The uniqueness of this prescription is spoiled in degenerate cases, but such states are of measure zero in the state space of any reasonable quantum system.  This is similar to the einselection criterion proposed by Zurek~\cite{Zurek1}, with one important distinction \---- Zurek's einselected basis of $Q$ is chosen by interactions with the environment which causes the state of $Q$ to undergo decoherence.  In our proposal, this basis (a basis of $A$, not of $Q$!) is selected entirely by the measurement interaction alone.  Thus the experimental setup selects the basis to be observed (which is in keeping with our intuition).

One might object to this proposal on the following grounds:  Suppose in the above experiment (performed by Alice with $Q$ a qubit for this example), the final state is \emph{not} given by the perfectly entangled state:
\begin{eqnarray}
|QA\rangle = \alpha|0,0\rangle + \beta|1,1\rangle
\end{eqnarray}
but in keeping with inefficient, realistic detectors, is instead given by:
\begin{eqnarray}
|QA\rangle = \alpha|0,0\rangle + \gamma|1,0\rangle + \delta|1,1\rangle. 
\end{eqnarray}
That is, there is an amplitude that Alice does not see $Q$'s transition to the $|1\rangle$-state, since the detector is not completely reliable.  In this case, $\rho_{A}$ describing Alice is no longer diagonal in the $\left\{|0\rangle, |1\rangle\right\}$ basis she set out to observe, yet Alice certainly expects to see one of these two outcomes, not a superposition of them.

This may at first appear to be a problem, since we do not perceive our experience basis to be dependent on the efficiency of the detectors we use.  However, what is failing here is just our oversimplification of the observer-detector as a single system, $A$.  In reality, the detector is a separate system $D$ which first becomes entangled with the quantum system $Q$, and then becomes entangled with the observer system, $A$.  Thus to handle this issue we must expand the description of the observed state further \---- the state of $QDA$ representing a less than perfect measurement is now expressed as: 
\begin{eqnarray}
|QDA\rangle = \alpha|0,0,0\rangle + \gamma|1,0,0\rangle + \delta|1,1,1\rangle.
\end{eqnarray}
Now the reduced density operator describing the observer Alice is again diagonal in the $\left\{|0\rangle, |1\rangle\right\}$ basis she set out to measure, and the probabilities again have their standard interpretation.  Note that we assumed again perfect entanglement between $D$ and $A$, but this is now a perfectly reasonable assumption, since these are macroscopically distinguishable states \---- i.e. there is never any real inefficiency in the detector's ability to communicate detection to the observer.  If the detector goes ``click," indicating a detection of the $|1\rangle$ state, we are safe in assuming that an attentive observer will always be able to distinguish this detector state from the absence of a click.  Of course there are many more steps analogous to this in a still more complete description of the detector and observer, but the central point is that as long as the \emph{final} communication between detector and observer can be regarded as reliable, then no basis problem exists in this formalism.  In the idealized limit of perfectly efficient detectors, however, we may safely bypass this additional step and consider the detector and observer to be a single system \---- we will often do this in what follows, suppressing the observer-detector distinction unless a detailed description of the detection interaction requires us to abandon the assumption of perfectly efficient detectors.

\subsection{The RR Covariant Quantum Theory}

The RR formulation of covariant quantum theory is somewhat different than that of orthodox quantum theory, but it is far more general~\cite{Rovelli1, Rovelli2, Rovelli3}.  To build an intuitive grasp for this formalism, we will first review the structure of orthodox quantum theory (in the Schr\"odinger picture), and then formulate covariant quantum theory for a direct comparison.  Much of this comparison can be found in more detail in Reference~\cite{Rovelli3}.

Orthodox quantum theory follows from the following postulates:

\begin{enumerate}
	\item States:  The possible states of a system are represented as vectors $\psi$ in a complex Hilbert space, $\mathcal H_0$.
	\item Observables:  Quantum observables are represented by Hermitian operators on $\mathcal H_0$.  The spectral theorem guarantees that any state of the system can be expanded in the basis of eigenstates of a complete set of commuting observables e.g. in the simple case of a single operator $A$, we can always write $|\psi \rangle = \alpha_{1} |a_{1} \rangle + \alpha_{2} |a_{2} \rangle + \alpha_{3} |a_{3} \rangle + ...$, where all the states $|a_{i} \rangle $ are eigenstates of $A$ with real eigenvalues $\lambda_{i}$.
	\item Evolution:  States evolve from one to another in an external, classical time parameter $t$ according to the Schr\"odinger equation, $ i \hbar \partial_{t} \psi(t) = H_{0} \psi(t)$.
	\item  Probability:  If we measure the observable $A$ of the system, the probability to observe the value $\lambda_{i}$ is given by $|\Pi_{i} |\psi \rangle|^2$, where $\Pi_{i}$ is the projector onto the eigenstate $|a_{i} \rangle$ with eigenvalue $\lambda_{i}$ (Born interpretation).  If outcome ``$i$" is observed, the system is ``collapsed" to the state $|a_{i} \rangle$ and will subsequently evolve from this state.
\end{enumerate}

In the previous subsection, we described a way to replace postulate (4) with a new postulate which preserves the unitary evolution implied by postulate (3), while giving consistent predictions.  We set this formalism aside in the present subsection (returning to it in the next), and consider the formulation of covariant quantum theory as presented by Reisenberger and Rovelli~\cite{Rovelli3}.

Before we state the postulates of covariant quantum theory, we define some terms and constructs that are not typically used in the traditional formulation of orthodox quantum theory.  These include: 
  
\begin{description}
	\item[Partial observables:]
	Partial observables are physical quantities which can be measured by an appropriate measuring apparatus and procedure.  The space of partial observables for a particular physical system is called its \emph{extended configuration space}, which we denote as $\mathcal M$.~\cite{Rovelli7}
\end{description}

\begin{description}
	\item[Complete observables:]
	Complete observables are relations between physical quantities that can be predicted from the knowledge of the state of a system.
\end{description}

Covariant language forces a careful distinction between ``partial observables," which are simply experimentally accessible quantities, and ``complete observables," which are quantities that represent the conjunction of two or more partial observables, and can be predicted from the physical state of a system.  For example, the position of a classical harmonic oscillator, $q$, is a partial observable.  The time, $t$, is also a partial observable.  These are both experimentally accessible quantities.  Note, however, that a series of results of $q$-measurements alone cannot be used to determine the current and future state of the oscillator, nor can a series of $t$-readings from a clock.  However, a series of measurements of $q(t)$, which simultaneously combine a $q$-measurement with a $t$-measurement, \emph{can} be used to reconstruct the current and future states of the oscillator (represented by a curve in the $q-t$ plane, i.e. a curve in $\mathcal M$) \---- thus, $q(t)$ represents a \emph{complete} observable.

These concepts underlie general covariance in a direct way.  In the context of general relativity, Einstein was led to diffeomorphism-invariant equations of motion by noting that coordinate points on the spacetime manifold have no objective meaning \---- in this language, they represent neither partial nor complete observables.  Instead, he noted that all predictions are made through the comparison of physical, dynamical quantities to other physical, dynamical quantities \---- a concept he referred to as ``spacetime coincidences."  In the present context, the points in $\mathcal M$ represent possible spacetime coincidences a given system might predict.  The principle of general covariance, then, is the statement that there is no universally preferred parameterization of structures on $\mathcal M$.  In simple mechanical systems involving few degrees of freedom (such as the systems we will consider in this paper), this implies that a theory must be general enough to express predictions independent of a preferred time coordinate on $\mathcal M$.  In more sophisticated systems like full GR or quantum gravity, it implies diffeomorphism invariance on the spacetime manifold.

\begin{description}
\item[Kinematical Hilbert space:]
  The space of $L^{2}$ functions on the extended configuration space, $\mathcal M$, with respect to an appropriate specified measure.  We denote the kinematical Hilbert space as $\mathcal K$.
\end{description}

\begin{description}
\item[Physical Projector:]
 The physical projector, $P$, takes arbitrary functions in $\mathcal K$ into solutions to the equations of motion, the space of which is denoted by $\mathcal H$.  That is:
 \begin{eqnarray}
P:\ \mathcal K & \rightarrow & \mathcal H \\
P:\	\psi^{\mathcal K}(x) & \mapsto & \psi^{\mathcal H}(x) \\
\psi^{\mathcal H}(x) & = & \int_{\mathcal M} dx' \ W(x;x') \psi^{\mathcal K}(x')  
\end{eqnarray}
Where $W(x;x')$ is the propagator for the theory, which is generally determined by the Hamiltonian~\cite{Rovelli1}.  
\end{description}

\begin{description}
\item[Physical Hilbert space:]
 This is the space of solutions to the equations of motion on $\mathcal M$, which generally have support over all of $\mathcal M$ (not compact support).  We denote this space as $\mathcal H$ (as above).  If each solution $\psi^{\mathcal H}$ in $\mathcal H$ can be obtained by projecting a state $\psi^{\mathcal K}$ of $\mathcal K$, then $\mathcal H$ can be made into a Hilbert space with the following inner product: \begin{eqnarray}
\langle \psi^{\mathcal H}| \phi^{\mathcal H} \rangle & = & \langle \psi^{\mathcal K} | P | \phi^{\mathcal K} \rangle \\
\langle \psi^{\mathcal K} | P | \phi^{\mathcal K} \rangle & = & \int_{\mathcal M} dx \ dx' \ \psi^{\ast \mathcal K}(x) W(x;x') \phi^{\mathcal K}(x')
\end{eqnarray}

We also have the property that $\langle x| P^{\dagger} P| x' \rangle = \langle x| P| x' \rangle =  W(x;x')$~\cite{Rovelli3}.  Typically, there are many kinematical states which project to the same physical state.
\end{description}

Covariant quantum theory can now be formulated according to the following postulates~\cite{Rovelli3}:

\begin{enumerate}
	\item States:  Kinematical states encode information on the outcome of measurements of experimentally accessible quantities (partial observables), and are represented as vectors in the Hilbert space $\mathcal K$.  Physical states encode information on the dynamical predictions of the theory \---- they are solutions to the equations of motion, $H \Psi = 0$ (the Wheeler-DeWitt equation, where $H$ is known as the relativistic Hamiltonian, or the Hamiltonian constraint).  These states are vectors in the Hilbert space $\mathcal H$.
	\item Observables:  Partial observables are represented by Hermitian operators on $\mathcal K$.  A given Hermitian operator on $\mathcal K$ also represents a complete observable if it additionally commutes with the relativistic Hamiltonian, $H$.  States of $\mathcal K$ are spanned by eigenstates of a complete set of partial observables.  States of $\mathcal H$ are spanned by eigenstates of a complete set of complete observables.
	\item Probability:  If the initial kinematical state can be described as $|\phi^{\mathcal K} \rangle$, then the probability of observing the state $|\psi^{\mathcal K} \rangle$ is given by the square of the \emph{physical} inner product, i.e.:
	\begin{eqnarray}
\mathcal P_{\phi \rightarrow \psi} & = & \left| \langle \psi^{\mathcal H} | \phi^{\mathcal H} \rangle \right|^{2} \\
 & = & \left| \langle \psi^{\mathcal K} | P | \phi^{\mathcal K} \rangle \right|^{2} \\
 & = & \left| \int_{\mathcal M} dx \ dx' \ \psi^{\ast \mathcal K}(x) W(x;x') \phi^{\mathcal K}(x') \right|^{2}
 \end{eqnarray}
After the measurement of a particular combination of partial observables corresponding to the state $|\psi \rangle$, the system is now described by the physical state $|\psi^{\mathcal H} \rangle = P |\psi^{\mathcal K} \rangle$.  We denote this as the RR probability postulate.
\end{enumerate}

Note that the ``evolution" postulate has in a sense become absorbed into the ``states" postulate, and the notion of observable has simultaneously been refined to accommodate this fact \---- in this way, explicit evolution in terms of an external, classical time variable is avoided as a fundamental ingredient of the theory, as required by generally covariant theories.  Covariant quantum theory is also designed to encompass traditional QM \---- it is merely designed to be more general.  To establish a bit of familiarity for this formalism, we examine a familiar physical system, quantized in this manner:

\begin{description}
\item[Single nonrelativistic particle:]
The extended configuration space $\mathcal M$ for a single particle confined to the real line is $\mathbb R^{2}$, which differs from the ordinary nonrelativistic configuration space ($\mathbb R$) because time must be included as a partial observable.  This is due to the fact that dynamical predictions generally require us to make some form of physical time measurement, and measurable quantities must correspond to Hermitian operators on $\mathcal M$.

The Kinematical state space $\mathcal K$ is thus $L^{2}(\mathbb R^{2})$, and is spanned by common eigenstates of $\hat{x}$ and $\hat{t}$ (representing partial observables), $|x,t \rangle$.  As is well known, this space has ``non-physical" properties, such as the absence of a ground state, due to the fact that the spectrum of $\hat{t}$ is the real line.  Note, however, that in this formalism, $\mathcal K$ is merely a space on which experimentally accessible quantities are defined as Hermitian operators \---- it does not contain information on which ``physical" states are allowed into the theory.

The relativistic Hamiltonian takes the form $H = H_{0} - i \hbar \frac{\partial}{\partial t}$, where $H_{0}$ is the familiar, non-relativistic Hamiltonian, and the equations of motion are then $H \Psi(x,t) = 0$.  Solutions to these equations of motion constitute the physical state space.

The measurement of $x$ and $t$ determines a localized state on $\mathcal K$ \---- for example, if an $\hat{x}$-measurement is localized around the point $x_{0}$ with precision $a$ and a $\hat{t}$-measurement is localized around the point $t_{0}$ with precision $b$, then (depending on how the measurement is actually performed), we might have the state $\phi^{\mathcal K}(x,t) = \exp \left[\frac{-(x-x_{0})^{2}}{a^{2}} + \frac{-(t-t_{0})^{2}}{b^{2}}\right]/(2 \pi ab)$ (which takes into account the fact that time measurements have a finite resolution, just as well as space measurements).

This then projects to a full physical state in $\mathcal H$, which represents the dynamics of the particle \emph{after} such a localizing measurement has been performed:
\begin{eqnarray}
\phi^{\mathcal H}(x,t) = \int dx' \ dt' \ W(x,t;x',t') \phi^{\mathcal K}(x',t')
\end{eqnarray}
where, for the free Schr\"odinger equation, $W(x,t;x',t') = \left( \frac{2 \pi m}{i \hbar (t - t')}\right)^{1/2} \exp \left( \frac{m (x - x')^2}{2 i \hbar (t - t') } \right)$~\cite{Rovelli1}.
\end{description}

The attractive quality of this formalism is, of course, its generality.  If instead of single-particle quantum mechanics we wished to study a system which truly requires a covariant treatment, for example minisuperspace cosmology with a scalar field, the picture would remain much the same.  The extended configuration space would again be $\mathbb R^{2}$ \---- the space of the variables $\Omega$ (related to the cosmological expansion parameter $a$ by $a = e^{\Omega}$), and $\phi$ (the average value of the scalar field).  Kinematical states (representing possible measurement outcomes or ``quantum coincidences") would again be $L^{2}(\mathbb R^{2})$, while the Hamiltonian takes the form
\begin{eqnarray}
H = \hbar^2 e^{-3 \Omega} \left( \frac{1}{24} \frac{\partial^{2}}{\partial \Omega^{2}} - \frac{1}{2} \frac{\partial^{2}}{\partial \phi^{2}} \right) - 6k e^{\Omega} + e^{3 \Omega} V(\phi) \nonumber,
\end{eqnarray}
from which $\mathcal H$ and the propagator are constructed~\cite{Isham1}.  Note that no time variable \emph{at all} appears in such a theory \---- there is not even an obvious variable that is \emph{analogous} to time (though often one variable is arbitrarily chosen to ``be" the time, in order to force the theory into the formalism of standard QM).  Nevertheless, the interpretation in terms of covariant quantum theory is essentially identical to that of single-particle quantum mechanics \---- time is no longer fundamental within this picture.

Though tremendously general, this formalism has had one critically weak link.  The RR measurement postulate (postulate 3) of covariant quantum theory (namely $\mathcal P_{\phi \rightarrow \psi} = \left| \langle \psi^{\mathcal K} | P | \phi^{\mathcal K} \rangle \right|^{2}$) does not reproduce the measurement postulate of orthodox quantum mechanics, though it is clearly constructed in analogy to the standard Born interpretation.  This was realized immediately by Reisenberger and Rovelli~\cite{Rovelli1}.  To obtain agreement with standard QM in the case of single measurements requires the use of a small-region approximation, which assumes that $\psi^{\mathcal K}$ has support in a sufficiently small region of the extended configuration space.  For multiple measurements, it originally appeared to require a pre-defined time variable to order the measurements, though Hellmann, Mondragon, Perez, and Rovelli have recently argued that any series of measurements can be treated mathematically within the formalism as a single measurement, for the purposes of making predictions.  We will elaborate upon these points further in the next section, where we will argue that the Reisenberger-Rovelli postulate cannot be a \emph{general} probability postulate.

\subsection{Covariant Quantum Information (CQI)}

Recently, we have proposed a quantum information theoretic interpretation of measurement within covariant quantum theory that appears to solve the above difficulties~\cite{Olson1}.  The idea can be thought of as an extension of the description of standard quantum measurement given by Cerf and Adami (CA), which we reviewed above.  That is, we include a model of the observer in any experimental scenario, and obtain probabilities by looking at the observer's reduced density operator \---- generally speaking, due to the properties of entanglement, quantum theory predicts that a measurement will result in an initially pure observer state evolving into a mixed one, and thus the predictions of quantum mechanics are inherently probabilistic. 

To invoke these ideas in the context of covariant quantum mechanics, we clearly need a partial trace.  This is simple on the kinematical Hilbert space $\mathcal K$, which has the usual $L^{2}$ tensor product form, but it is not so obvious on the physical Hilbert space $\mathcal H$ in which states cannot generally be expressed as a tensor product of subsystem states.

To begin, we define a covariant system analogous to $Q$ by specifying its extended configuration space $\mathcal M_{Q}$ and a relativistic Hamiltonian $H_{Q}$.  To include an observer system analogous to $A$, we enlarge the configuration space via the Cartesian product, i.e. $\mathcal M = \mathcal M_{Q} \times \mathcal M_{A}$, and define a new Hamiltonian $H$ for the combined system.  In what follows, let $x$ represent coordinates of $\mathcal M_{Q}$, and let $y$ represent coordinates of $\mathcal M_{A}$ \---- i.e. we express a general point on $\mathcal M$ with the pair, $\left\{x,y\right\}$. 

By analogy with Cerf and Adami, we wish the partial trace to express local information held by a specific subsystem, but not by using a preferred time variable.  Instead, we select a generic region of interest $\mathcal S$ somewhere in $\mathcal M$.  In covariant theories like general relativity, there is in general no diffeomorphism-invariant meaning of a general region on the \emph{spacetime} manifold, since spacetime coordinates represent neither partial nor complete observables \---- in the present case, however, we are working directly on the space of experimentally observable quantities $\mathcal M$, thus our regions $\mathcal S$ have a definite meaning in terms of experimental data.  In the limit of the Schr\"odinger picture, the region $\mathcal S$ corresponds to a constant time slice of space, but in general $\mathcal S$ may be chosen freely, provided a few conditions are met:  We require that the physical state $\phi^{\mathcal H}$ under consideration can be expressed via the projection of a $\mathcal K$ state $\phi^{\mathcal K}$ with support in $\mathcal S$ \---- this requirement simply means that the physical state of the system can be reconstructed from local experimental data (also, this local state is normalized with respect to $\mathcal H$, i.e. we require $\langle \phi^{\mathcal K}| P | \phi^{\mathcal K} \rangle = 1$).  Next, we require that for all points $\left\{x,y\right\}$ and $\left\{x',y'\right\}$ in $\mathcal S$, the propagator takes the form $W(x,y;x',y') = W_{Q}(x;x')\delta(y - y')$, where $W_{Q}(x;x')$ is obtained from the free Hamiltonian $H_{Q}$.  This expresses the fact that we are considering a region $\mathcal S$ where no interactions between system $Q$ and $A$ are taking place (e.g. before or after the measurement interaction has taken place) and the evolution of system $A$ is trivial.  That is, if a measurement has been performed, the measuring system $A$ does not forget about the outcome anywhere in $\mathcal S$ \---- it merely holds the relevant information.  Such a region $\mathcal S$ satisfying these conditions is a generalized replacement for the concept of a constant-time slice (which satisfy these requirements in standard QM).

When these conditions are met, we can approximate the full Wheeler-DeWitt equation $H(x,y)\Psi(x,y)=0$ as $H_{Q}(x)\Psi(x)=0$ in our region of interest, and thus the physical state space $\mathcal H$ is locally indistinguishable from $\mathcal H_{Q} \otimes L^{2}(\mathcal M_{A}) \equiv \mathcal H_{Q} \otimes \mathcal K_{A}$.

We define a new projector, $P_{Q}$, from $\mathcal K_{Q}$ to $\mathcal H_{Q}$, so that we can now express our state as a physical density operator via $\rho = P_{Q} |\phi^{\mathcal K} \rangle \langle \phi^{\mathcal K}| P_{Q}^{\dagger}$ (valid only in the region $\mathcal S$).  Now express $|\phi^{\mathcal K} \rangle$ via a Schmidt decomposition as $\sum_{i} \lambda_{i} |\phi^{\mathcal K_{Q}}_{i} \rangle |\phi^{\mathcal K_{A}}_{i} \rangle$.  $P_{Q}$ operates only on $\mathcal K_{Q}$, so we can take a partial trace over $\mathcal K_{Q}$ and using the cyclic property of the trace, the properties of the projector, and the physical inner product ($\langle x| P^{\dagger} P| x' \rangle = W(x;x')$), we obtain a reduced density operator on $\mathcal K_{A} \equiv L^{2}({\mathcal M_{A}) }$:
\begin{widetext}
\begin{eqnarray} 
	\rho_{A} = \sum_{i,j} \int_{\mathcal M_{Q}} dx\ dx'\ \lambda_{i} \lambda_{j} \phi^{\mathcal K_{Q}}_{i}(x) W_{Q}(x;x') \phi^{\ast \mathcal K_{Q}}_{j}(x') |\phi^{\mathcal K_{A}}_{i} \rangle \langle \phi^{\mathcal K_{A}}_{j}|.
\end{eqnarray}
\end{widetext}
This is now a reduced density operator on $\mathcal K_{A}$, containing the physically relevant information locally available to the observer $A$ within $\mathcal S$.  Note the range of integration is contained entirely within $\mathcal S$ due to the support of the $L^{2}$ functions $\phi^{\mathcal K}$.  This covariant definition immediately reduces to the ordinary definition of the partial trace when the region of interest $\mathcal S$ is a constant time slice as before, but it is clearly more general \---- $\phi^{\mathcal K}$ may be smeared in any number of ways over a non-zero time interval (if there is a time variable), provided that the observer system $A$ is making no transitions in $\mathcal S$.  It also remains meaningful for fully covariant systems having no pre-defined time variable at all.

With this partial trace we have implemented the idea of Cerf and Adami covariantly, without any preferred coordinates on $\mathcal M$, and without any external time variable.  Within a given region $\mathcal S$, an observer will be described by a mixture on his kinematical state space $\mathcal K$, even if we know the full physical state in $\mathcal H$ is pure.  The probabilities and experience basis can be extracted exactly as we have described in subsection A \---- by finding the unique diagonal, orthogonal basis of $\rho_{A}$ obtained from the covariant partial trace (note that although the Schmidt decomposition in $\mathcal K_{Q} \otimes \mathcal K_{A}$ defines a basis in which the reduced density operator obtained from $\phi^{\mathcal K}$ is diagonal when the partial trace is over $\mathcal K_{Q}$, the covariant partial trace does not necessarily leave $\rho_{A}$ diagonal in this basis).

Our covariant approach to quantum information has several important features which we mention here, and it is the purpose of the sections that follow to elaborate them in more detail.

\begin{itemize}
	\item \textbf{CQI agrees with standard QM:}  Due to the fact that the covariant partial trace reduces exactly to the standard definition when the region $S$ corresponds to a constant time slice of space, the formalism of Cerf and Adami is guaranteed to emerge in the limit of standard Schr\"odinger equation dynamics and the usual (though unphysical) assumption of arbitrarily good clocks.  This will be demonstrated in the next section, where we show how correspondence with the standard Born interpretation is recovered in the context of CQI.  The original objections to the RR picture of covariant quantum theory based on its failure to correspond with standard QM predictions are resolved in the context of CQI.
	
	\item \textbf{CQI is compatible with realism:}  Remarkably, all predictions for all observers can be obtained from a single physical state in the theory, implying that a single, objective state could in principle be used to describe the universe, rather than the essential use of different states for different observers.  We will demonstrate this feature in more detail in sections IV and V.
	
	\item \textbf{CQI is fully unitary:}  Because there is no collapse or reduction of the full state vector upon observation, all evolution in the theory is unitary.  This means that time reversal is a good symmetry, even in the presence of quantum measurements.  This will allow us to make new experimental predictions for phenomena associated with measurement, e.g. the quantum Zeno effect, which we will consider in section VI.
	
	\item \textbf{CQI is fully covariant:}  By construction, our formalism is compatible with generally covariant physics.  The theory is formulated directly on the space of physically accessible quantities $\mathcal M$ (the points of which correspond to the ``spacetime coincidences" described by Einstein in the context of classical GR).  No parameterization of the states on $\mathcal M$ is pre-selected to play any special role to formulate the theory or to obtain predictions from it (though it is clear that in many cases the dynamics under investigation may make one choice of coordinates far more \emph{convenient} than another, as is the case with single-particle Schr\"odinger dynamics).
	
	\item \textbf{CQI is fully self-contained:}  Since all observers are modeled quantum mechanically, there is no sense in which probabilities are assumed to refer to an external, classical agent.  This property, which it shares with the original Everett-interpretation of quantum measurement~\cite{Everett1}, makes the formalism natural for quantum cosmology, where the concept of an external observer is undefined, and debate has centered on the interpretation of quantum probabilities for this reason.  Section VII will comment on this feature.
\end{itemize}  

Thus, covariant quantum information appears to cast new light on foundational issues of quantum mechanics and quantum cosmology, and simultaneously allows us to make use of generally covariant quantum mechanics for all quantum systems, in essence establishing a working connection between concrete atomic-scale quantum measurements, and those of untested Planck-scale theories of quantum gravity.

\section{Comparison of probability postulates for covariant quantum theory}

It is important to realize that our covariant quantum information framework predicts \emph{different} probabilities than does the Reisenberger-Rovelli postulate.  This is because entanglement is a necessary feature of measurement in our covariant information approach, whereas the Reisenberger-Rovelli probability postulate assigns probabilities between any two kinematical states.  Thus, there are many situations in which the Reisenberger-Rovelli postulate assigns a probability but our framework does not \---- for example, whenever we ask about the probability that a system is found in state $\psi^{\mathcal K}$, without specifying a quantum system to perform the measurement and hold information on its outcome.

Equally important is that this shift in interpretation can lead to different probabilities for the same physical question.  To see this explicitly, we analyze here single-particle dynamics, where we can make contact with the well-known Born interpretation as a correspondence rule that must be reproduced in the appropriate limit.

It was immediately clear to Reisenberger and Rovelli that there was a difficulty with their original probability postulate~\cite{Rovelli1, Rovelli2, Rovelli4, Rovelli6}.  Here we consider the system described in their original paper~\cite{Rovelli1} \---- a single particle satisfying the Schr\"odinger equation, and a simple two-state detector designed to activate if the particle passes through spacetime region $\mathcal R$.  The detector is prepared in state $|0 \rangle$, and transitions to state $|1 \rangle$ if the particle interacts with the detector in $\mathcal R$.  The interaction Hamiltonian can be written as $H_{\mathrm int}(x) = \alpha V(x) (|1\rangle \langle 0| + |0\rangle \langle 1|)$, where the potential $V(x)$ is constant in $\mathcal R$ and zero elsewhere.  The potential is assumed to be sufficiently weak so that first order perturbation theory can be used, and thus transitions from the state $|1\rangle$ back to the state $|0\rangle$ can be ignored.  In the following analysis, the coordinates $x = \left\{ X, T \right\}$ are used.

To ask the question ``what is the probability that the detector is activated?" in the standard Born-rule interpretation, we need an initial state $\Psi_{0}(x)$ at some initial time $T_{0}$ and we must specify some later time $T$ after the interaction to look at the form of the wave function. Reisenberger and Rovelli showed that at late time $T$ the combined system has evolved to a state of the form:
\begin{widetext}
\begin{eqnarray}
\int dX'\ W(X,T;X',T_{0}) \Psi_{0}(X',T_{0})\ |0\rangle +  \frac{\alpha}{i \hbar}  \int_{\mathcal R} dX'\ dT'\ W(X,T;X',T') V(X',T') \Psi(X',T')\ |1\rangle,
\end{eqnarray}
\end{widetext}
where here $\Psi(x) = \int dx'\ W(x,x') \Psi_{0}(x')$.  From here, the Born rule tells us that the probability that the detector was activated is the square of the second term integrated over $X$ at the constant time $T$.  With some manipulation and use of the properties of the Schr\"odinger propagator, it can be shown that this probability takes the following form:
\begin{eqnarray}
\mathcal P_{\mathrm Born} = \frac{\alpha^{2}}{\hbar^{2}} \int_{\mathcal R} dx \int_{\mathcal R} dx' \Psi^{\ast}(x)W(x;x')\Psi(x').
\end{eqnarray}
Now, for this scenario, the postulate of Reisenberger and Rovelli for covariant quantum theory is that:
\begin{eqnarray}
\mathcal P_{\mathrm R- \mathrm R} &=& |\langle \mathcal R |P| \Psi^{\mathcal K}_{0} \rangle |^{2} \\
&=& \left| \int dx\: \int dx'\:  \mathcal R^{\ast}(x) W(x;x') \Psi^{\mathcal K}_{0}(x') \right|^{2}.
\end{eqnarray}
Here, $\mathcal R(x)$ is a normalized, constant function in $\mathcal R$, and zero elsewhere.  $\Psi^{\mathcal K}_{0}$ is the initial state prepared on a constant-time slice as above.  Using the definition of $\Psi(x)$ above, we can re-write this as follows:
\begin{eqnarray}
\mathcal P_{\mathrm R- \mathrm R} &=& \left| \int dx\: \mathcal R^{\ast}(x) \Psi(x) \right|^{2} \\
&=& \int dx\: \int dx'\ \Psi^{\ast}(x) \mathcal R^{\ast}(x) \mathcal R(x') \Psi(x') \\
& = & \int_{\mathcal R} dx\: \int_{\mathcal R} dx'\ \Psi^{\ast}(x) \Psi(x').
\end{eqnarray}
This form of $\mathcal P_{\mathrm R- \mathrm R}$ is similar to the form of $\mathcal P_{\mathrm Born}$ shown in equation (24), except for the missing factor of the propagator in the integrand.  If $\mathcal R$ is taken to be sufficiently small, then these two results can be made to agree up to a constant factor related to the description of the measurement apparatus.  A similar small-$\mathcal R$ limit has been taken in all previous presentations of RR covariant quantum theory.  However, it is easy to construct examples outside of this limit where disagreement immediately arises.  For example, take $\mathcal R$ to be the union of two points at $x=a$ and $x=b$ on a single constant-time slice (so that $W(a;b)\approx 0$) \---- then we have that $\mathcal P_{\mathrm Born} \propto |\Psi(a)|^{2} + |\Psi(b)|^{2}$, but that $\mathcal P_{\mathrm R- \mathrm R} \propto |\Psi(a)|^{2} + |\Psi(b)|^{2} + \Psi^{\ast}(a)\Psi(b) + \Psi^{\ast}(b)\Psi(a)$.  If $a$ and $b$ are sufficiently separated, then the probabilities cannot be made proportional.

This illustrates an extremely important point \---- if the original Reisenberger-Rovelli probability postulate fails to reproduce what we already know to be true in standard atomic-scale quantum mechanics, it can certainly not be taken for granted in the unexplored Planck-scale or entire universe-scale regimes, for which it was designed to be used in the first place.

Our covariant quantum information approach, by contrast, has no difficulty here.  Recall that our CQI framework requires us to define a region of the extended configuration space, $\mathcal S$, in order to obtain probabilities.  If we take the same measurement apparatus defined above, and if $\mathcal S$ is taken to be a constant-time slice of space, then the Born result is reproduced identically.  However, we are by no means \emph{required} to take $\mathcal S$ to be a constant-time slice \---- one main motivation for this approach to physics was in fact to do away with the preferred status of the time variable in the construction of the theory.

To see this explicitly in this example, note that to the future of the interaction region $\mathcal R$, the detector is evolving trivially, thus the physical Hilbert space $\mathcal H$ is indistinguishable from $\mathcal H_{Q} \otimes \mathbb C^{2}$ in this region \footnote{Actually, by the argument of section IIA, we take the system to be $\mathcal H_{Q} \otimes \mathbb C^{2} \otimes \mathbb C^{2}$ in this region, making a distinction between the detector $D$ and the observer $A$, which become entangled to the future of $\mathcal R$.  We then take the covariant partial trace over both $\mathcal H_{Q}$ and $D$.  We are suppressing the explicit mention of the additional detector system in the calculation to avoid distraction based on this finer point of the formalism, which is required to do away with off-diagonal elements in the reduced density operator in the $|0\rangle, |1\rangle$ basis.}.  Thus all regions to the future of $\mathcal R$ are equivalent for the purposes of calculating the transition probability, so long as they are large enough to support a $\mathcal K$ state $\Psi^{\mathcal K}(x)$ that projects to the relevant physical states on $\mathcal H$.  Using the covariant partial trace, the detector's reduced density operator can be expressed as follows:
\begin{eqnarray}
\rho_{A} =  \left[\frac{\alpha^{2}}{\hbar^{2}} \langle \phi | \phi \rangle^{\mathcal H_{Q}}\ |1 \rangle \langle 1| + \langle \psi | \psi \rangle^{\mathcal H_{Q}}\ |0\rangle \langle 0 | \right]
\end{eqnarray}
where $\phi(x)$ and $\psi(x)$ are defined as follows:
\begin{eqnarray}
\phi(x) &=& \int_{\mathcal R} dx' \int_{\mathcal M_{Q}} dx'' W(x;x') V(x') W(x',x'') \Psi_{0}(x'') \nonumber \\
\psi(x) &=& \int_{\mathcal M_{Q}} dx'\ W(x;x') \Psi_{0}(x'),
\end{eqnarray}
which represent unnormalized states in $\mathcal H_{Q}$.  The probability for the detector to be activated is thus given by:
\begin{eqnarray}
\mathcal P_{\mathrm CQI} = \frac{\alpha^{2}}{\hbar^{2}} \langle \phi | \phi \rangle^{\mathcal H_{Q}},
\end{eqnarray}
which depends on the physical inner product in $\mathcal H_{Q}$ rather than on the preferred status of any time coordinate.  However, because of the isomorphism between $\mathcal H_{Q}$ and $L^{2}(\mathbb R)$ on a constant time slice~\cite{Rovelli3}, we can immediately see that this probability reduces to that obtained by the Born interpretation above, which \emph{does} make use of a preferred time coordinate (but is well-established experimentally).  Using the above definitions of $\phi(x)$ and $\Psi(x)$, we have:
\begin{widetext}
\begin{eqnarray}
\mathcal P_{\mathrm CQI} &=&  \frac{\alpha^{2}}{\hbar^{2}} \langle \phi | \phi \rangle^{\mathcal H_{Q}}\\
&=&  \frac{\alpha^{2}}{\hbar^{2}} \langle \phi^{\mathcal K}_{t=T} |P| \phi^{\mathcal K}_{t=T} \rangle^{\mathcal K_{Q}}\\
&=&  \frac{\alpha^{2}}{\hbar^{2}} \int dX\  \phi^{\ast}(X,T) \phi(X,T) \\
&=&  \int dX\ \left|\frac{\alpha}{i \hbar}  \int_{\mathcal R} dX'\ dT'\ W(X,T;X',T') V(X',T') \Psi(X',T')\right|^{2}\\
&=&  \frac{\alpha^{2}}{\hbar^{2}} \int_{\mathcal R} dx \int_{\mathcal R} dx' \Psi^{\ast}(x)W(x;x')\Psi(x'),
\end{eqnarray}
\end{widetext}
which, though defined in a coordinate-independent, covariant manner by equation (32), reduces to exact agreement with the Born rule of equation (24) \footnote{There is a potentially confusing point here and it is important that it not be misunderstood.  Equation (32) is obtained from a generic region $\mathcal S$ to the future of the measurement interaction region $\mathcal R$, and does not depend upon the identification of any specific time variable.  However, in evaluating equation (32), we have made use of a constant-time slice on $\mathcal M_{Q}$.  The latter is merely a calculational convenience offered by the specific form of the Schr\"odinger propagator for the purposes of evaluating the physical inner product \---- it introduces no interpretation as a ``probability at time $T$," which is precisely what we are attempting to avoid.  The resulting probability is valid in any sufficiently large region $\mathcal S$ to the future of the measurement interaction, including those regions which might not even contain a constant-$T$ surface themselves.  The thing that is covariant about this treatment is that equation (32) is obtainable from a generic region $\mathcal S$ and does not depend on the identification of any time variable.  The means by which a given physical inner product might be evaluated, however, is not something that we necessarily need to banish ``$T$" from \---- we are free, of course, to take advantage of any convenient mathematical properties a given propagator might have in a particular coordinate system.}.  The entanglement properties of measurement and use of the covariant partial trace are responsible for the discrepancy with the Reisenberger-Rovelli postulate, and agreement with standard QM.  In our CQI framework, entanglement is \emph{essential} for obtaining probabilities, for it is only through entanglement that a partial trace leads a quantum observer into a mixed state.  This is not true of the R-R postulate, which assigns probabilities between any two states in $\mathcal K$.

Another difficulty associated with the Reisenberger-Rovelli probability postulate is that there is no preferred time variable or causal structure available to order multiple measurements.  For example, suppose we wish to measure the state $|\phi \rangle$, and also the state $|\xi \rangle$.  What is the probability that the system is measured in \emph{both} of these states?  In general, if the projector onto $|\phi \rangle$, $\Pi_{\phi}$, does not commute with the projector onto $|\xi \rangle$, $\Pi_{\xi}$, then we need to distinguish between $\left| \Pi_{\phi} \Pi_{\xi} |\Psi \rangle \right|^{2}$ and $\left| \Pi_{\xi} \Pi_{\phi} |\Psi \rangle \right|^{2}$, which will be two distinctly different probabilities.  In standard QM, this distinction is easy \---- we simply order the projections according to the time variable $t$ (or order them according to the causal structure on a fixed background).  However, a fully covariant, background-free formalism does not have such constructs \---- the theory must not be required to reference a preferred time coordinate of $\mathcal M$ in order to make predictions (likewise, nonperturbative quantum gravity must not be required to reference a fixed background causal structure to make predictions).

Recently, Hellmann, Mondragon, Perez, and Rovelli have argued that this time ordering problem can be avoided by interpreting all predictions within the RR covariant quantum theory strictly as single-measurement probabilities~\cite{Rovelli4, Rovelli6}.  That is, instead of constructing $\Pi_{\phi}$ and $\Pi_{\xi}$ separately, we enlarge the system to include at least one measuring apparatus that determines for example whether or not the state $\phi$ was observed.  We then take the projector onto the state $|\xi \rangle$ as well as the projector onto the state $|\phi = yes \rangle$ of the detector (call it $\Omega_{\phi}$).  Then the single projector $\Pi_{\xi}\Omega_{\phi}$ on this larger, combined system gives an effective multiple-measurement probability $\left| \Pi_{\xi}\Omega_{\phi} |\Psi \rangle \right|^{2}$ according to the original probability postulate, but makes use of only a single non-unitary collapse and so evades the multiple-measurement probability problem.  Note that in the case of $N$ measurements, this approach makes use of $N-1$ measurement devices to remove the ordering ambiguity.

This approach is similar to our own, which makes use of $N$ observers for $N$ measurements.  Note that in our strategy, no external time ordering is required because we simply specify a given region $\mathcal S$, and ask if a given measurement apparatus $A$, $B$, etc. is in a mixed state within this region.  If so, we compute probabilities exactly as before, for each measurement apparatus.  Because this formalism reduces identically to that of Cerf and Adami when we study simple Schr\"odinger-equation systems and constant-time slices, all the effective collapse features of the CA formalism are retained.  No collapse postulate is required, and thus no special time variable needs to be selected to order projections for multiple measurements.

The CQI formalism does, however, allow a different kind of time ordering to emerge in the context of multiple measurements.  We can pick out a series of regions $\mathcal S$, $\mathcal S'$, $\mathcal S''$, etc. and order them according to the von Neumann entropy of the observed system $Q$.  In the case of systems described by the Schr\"odinger equation, the regions corresponding to constant-time slices, and all measurement interactions corresponding to ideal projective measurements of $Q$, this ordering corresponds exactly to the ordering of the external $t$-variable, since it is known that projective measurements can only increase the entropy of a system~\cite{Nielsen1}.  It remains to be seen how closely this type of entropic time ordering resembles other notions of time in generally covariant systems.

To sum up, Hellmann, Mondragon, Perez, and Rovelli seek to solve the time-ordering problem by postponing a non-unitary collapse until the end of the analysis, while we \emph{never} make use of projective collapse within our CQI approach.  In addition, we resolve the Born correspondence problem in the case of single measurements, and define an entropic quantum mechanical arrow of time independent of any coordinate on $\mathcal M$ or background causal structure.  We do not interpret quantum probabilities as probabilities for the full physical state $\Psi^{\mathcal H}$ of the universe to make any kind of jump to a new physical state \---- the probabilities in covariant quantum information simply represent the ignorance inherent in the state of an entangled, localized observer.  In essence, our CQI picture realizes perfect correspondence with standard QM in the appropriate limit and removes the most serious objections to covariant quantum theory as proposed by Reisenberger and Rovelli.

\section{Realism}
Although the aim of covariant quantum information is primarily to remove ambiguities and to obtain correspondence with the well-established predictions of standard quantum mechanics in the appropriate limit, it also inherits a realist picture from the Cerf-Adami description of measurement~\cite{Adami4}.  It is remarkable that a well-known interpretational issue in the foundations of quantum mechanics is forced to the surface by our desire to express the concept of measurement covariantly \---- and even more remarkable, that a realist interpretation of the quantum state function $\Psi$ should naturally emerge in the process.

Our definition of quantum realism is the following:  We describe a quantum measurement formalism as \emph{realist} if at each individual time (or in the covariant language, each sufficiently large region of $\mathcal M$), all observers can correctly obtain all physical predictions about a quantum system from the same state function $\Psi$.  This definition allows for the possibility that the universe is described by a single quantum state in $\mathcal H$, independent of our knowledge of it.  It is straightforward to show that the standard description of projective measurement is not a realist formalism, according to this definition \footnote{Note that this conception of realism is distinct from that of ``Einstein realism" or ``local realism" which appear to be fundamentally at odds with observation, since the experimental violation of Bell's inequalities.  Here, we are primarily interested in whether or not there exists a quantum description of reality with which all observers can simultaneously agree and obtain ideal predictive power.}.

The prototype scenario to demonstrate this has been called ``the observer observed" by Rovelli~\cite{Rovelli3, Rovelli5}.  Here we start with a quantum system $Q$ in a superposition state $|Q\rangle = \alpha|0\rangle + \beta|1\rangle$, and two observers $A$ (Alice) and $B$ (Bob).  At time $t_{1}$, Bob performs a measurement of $Q$, in the $|0\rangle, |1\rangle$ basis.

We now ask the question, ``what is the state of the system?"  As described by Bob (using the standard formalism of projective measurement), the state of $Q$ has been collapsed and will either be described by $|0\rangle$ or $|1\rangle$, depending on the outcome of his measurement.  Alice, on the other hand, describes the system differently \---- no collapse has taken place, but the system has grown into an entangled state that encompasses Bob as well as $Q$, $|QB\rangle = \alpha|0\rangle|\mathrm{Bob\ sees}\ 0\rangle + \beta|1\rangle|\mathrm{Bob\ sees}\ 1\rangle$.

This necessary disagreement on the state used to describe the system $Q$ has led many thinkers to the conclusion that quantum states have no independent meaning outside of their use by a particular observer.  Quantum mechanical states are to be thought of a kind of book-keeping device which may be different for each separate observer.  Peres wrote, for example~\cite{Peres2}:
\begin{quotation}
In summary, the question raised by EPR ``Can quantum-mechanical description of physical reality be considered complete?" has a positive answer. However, reality may be different for different observers.
\end{quotation}
\ldots and Rovelli~\cite{Rovelli5}:
\begin{quotation}
If different observers give different accounts of the same sequence of events, then each quantum mechanical description has to be understood as relative to a particular observer. Thus, a quantum mechanical description of a certain system (state and/or values of physical quantities) cannot be taken as an ``absolute" (observer independent) description of reality, but rather as a formalization, or codification, of properties of a system relative to a given observer.
\end{quotation}

And indeed, these are quite reasonable conclusions if one takes the process of measurement to physically correspond to a projection onto states of $Q$.  However, we need not do this.  Remarkably, the principle that takes us out of this picture is precisely the same principle we used to remove the time ordering ambiguity for generally covariant quantum mechanics, and to obtain correspondence with the Born rule.  \emph{Predictions must be obtained from a quantum model of the observer.}
  
When we do this, we find that we have a new description of the observer observed.  Both Alice and Bob can agree that the initial state (a $\mathcal K$-state on hypersurface $t=t_{0}$) of the system is:
\begin{widetext}
\begin{eqnarray}
|QAB\rangle = (\alpha|0\rangle + \beta|1\rangle)|\mathrm{Alice\ ready} \rangle |\mathrm{Bob\ ready} \rangle.
\end{eqnarray}
On hypersurface $t=t_{1}$, after Bob performs his measurement, the system is described by:
\begin{eqnarray}
|QAB\rangle = (\alpha|0\rangle|\mathrm{Bob\ sees}\ 0\rangle + \beta|1\rangle|\mathrm{Bob\ sees}\ 1\rangle)|\mathrm{Alice\ ready} \rangle.
\end{eqnarray}
Predictions are obtained by analyzing Alice and Bob.  On this $t=t_{1}$ slice, we have for Alice that
\begin{eqnarray}
\rho_{A} = |\mathrm{Alice\ ready} \rangle \langle \mathrm{Alice\ ready}|
\end{eqnarray}
indicating that no outcome observed by Alice is yet probabilistic.  On the other hand, on the same hypersurface: 
\begin{eqnarray}
\rho_{B} = |\alpha|^{2}|\mathrm{Bob\ sees}\ 0 \rangle \langle \mathrm{Bob\ sees}\ 0| + |\beta|^{2}|\mathrm{Bob\ sees}\ 1 \rangle \langle \mathrm{Bob\ sees}\ 1|
\end{eqnarray}
(the unique diagonal, orthonormal representation of $\rho_{B}$) indicates that the outcome of Bob's experiment will be $|0\rangle$ or $|1\rangle$ with probabilities $|\alpha|^{2}$ and $|\beta|^{2}$, respectively.  

Now suppose that at $t=t_{2}$, Alice discusses with Bob the results of his experiment or she performs her own measurement on $Q$.  The state on this hypersurface is: 
\begin{eqnarray}
|QAB\rangle = \alpha|0\rangle|\mathrm{Bob\ sees}\ 0\rangle|\mathrm{Alice\ sees}\ 0\rangle + \beta|1\rangle|\mathrm{Bob\ sees}\ 1\rangle|\mathrm{Alice\ sees}\ 1 \rangle.
\end{eqnarray}
\end{widetext}
Like Bob, Alice will see outcomes $|0\rangle$ and $|1\rangle$ with respective probabilities $|\alpha|^{2}$ and $|\beta|^{2}$, but importantly, the \emph{conditional entropy} between Alice and Bob on this hypersurface is zero:  $S(A|B) = S(AB) - S(B) = 0$ \---- i.e. the outcome of Bob's experiment (which was performed at $t_{1}$) completely determines what Alice will see at time $t_{2}$. 

Thus, all predictions (including those usually ascribed to a collapse) are obtained from a \emph{single} physical state in $\mathcal H$, analyzed on different regions (in this case, constant time slices) of $\mathcal M$.  At no point were Alice and Bob forced to disagree on the physical state describing their situation, yet they are both able to make all of the standard predictions about what they will see.  Surprisingly, in the pursuit of consistently incorporating general covariance into a formalism of quantum measurement, we have been led directly into a working realist interpretation of quantum theory.  This paints an intuitive picture of quantum reality \---- that a single physical state $\Psi$ describes the universe; the past, the present, and the future.  We may not have complete knowledge of $\Psi$ (and in fact observers may typically disagree about the state of the system simply due to lack of information, just as they do in classical mechanics), but all probabilistic predictions for all observers at all times are ultimately, in principle, obtainable from it.

\section{Quantum collapse and EPR states}
Not surprisingly, a formalism that dispenses with collapse will have implications for non-locality in the EPR scenario~\cite{Einstein1}.  Quickly setting up the scenario, we have Alice located at spatial point $a$, and Bob at point $b$.  An entangled quantum state is prepared, $\alpha|00\rangle + \beta|11\rangle$, and Alice is given control of the first qubit, and Bob the second.  For simplicity here, we take the qubits to be abstract two-state systems that transform trivially under Lorentz transformations, i.e. the two-state system should not be thought of as a spinor in 3-space.

Suppose now that Alice performs a measurement on her qubit in the $\left\{|0\rangle, |1\rangle \right\}$ basis.  The standard explanation is that the full, nonlocal state has \emph{instantly} been collapsed to $|0\rangle|0\rangle$ or to $|1\rangle|1\rangle$.  Worse still, if Bob performs a measurement at an instant that is spacelike separated from Alice's measurement, there are frames of reference in which the order of collapse changes.  Some involve Alice collapsing the state, and Bob performing a measurement on a collapsed state, and some involve Bob collapsing the state.  This state of affairs has caused concern among many who notice the seeming contradiction with special relativity.  Happily for relativists, a no-communication theorem ensures that the effects of this collapse cannot be used to send superluminal information between Alice and Bob~\cite{Peres1}.

This is not the end of the story, though.  What did we mean when we said the state was collapsed \emph{instantly} by Alice?  Instantly in which frame of reference?  One immediate and seemingly reasonable answer is Alice's frame of rest~\cite{Suarez1}.  However, suppose a third observer, Charlie, were to pass by Alice with some velocity exactly as she performs her measurement.  If the non-local collapse is instant in Alice's frame, it cannot be instant in Charlie's frame, though he observed the measurement outcome at precisely the same moment as Alice.  Although this scenario is closely related to the issue of realism described in the previous section, the addition of relativity considerations to the mix causes thinking along these lines to be thorny enough that clever experiments have actually been performed to observe the speed of collapse of separated, entangled pairs~\cite{Gisin1} (incidentally, setting lower bounds of order $10^{4}c$ on the speed of quantum information). 

In our CQI picture, the description is clarified.  The wave function does not undergo a collapse, instantaneous or otherwise.  Probabilities are interpreted as the classical uncertainty of the observer's reduced state, which becomes a mixture via local entangling interactions with the EPR pair \---- we do \emph{not} have probabilities for any kind of non-local quantum operation, and require no propagation of any effect.  Alice and Charlie do not disagree on the way in which the state of the universe is collapsed, since there are simply no nonlocal effects happening at all.  The ``speed of collapse" is an undefined concept in this picture.  Alice and Charlie are both described by mixtures to the future of Alice's measurement event, and Bob is described by a mixture to the future of his own measurement event \---- these are entirely local statements, in which there is absolutely no contradiction.

The addition of relativity simply means that there are more spacelike surfaces (related by Lorentz transformations) on which we can express the full physical state in $\mathcal H$ as the projection of a kinematical state.  On some of these surfaces, Alice has performed a measurement (and is described by a mixture) but Bob has not.  On others, Bob has performed a measurement (and is described by a mixture), but Alice has not.  However, on \emph{any} constant-time slice in which both Alice and Bob could have compared notes via transmission of their measurement results, the $\mathcal K$-state describes Alice as entangled with Bob:
\begin{widetext}
\begin{eqnarray}
|Q_{1}AQ_{2}B\rangle = \alpha|0\rangle|\mathrm{Alice\ sees}\ 0 \rangle|0\rangle|\mathrm{Bob\ sees}\ 0 \rangle + \beta |1\rangle|\mathrm{Alice\ sees}\ 1 \rangle|1\rangle|\mathrm{Bob\ sees}\ 1 \rangle,
\end{eqnarray}
Calculating the reduced density operators for Alice and Bob, standard results are obtained.  Individually for Alice and Bob, we have:
\begin{eqnarray}
\rho_{A} &=& |\alpha|^{2} |\mathrm{Alice\ sees}\ 0\rangle \langle \mathrm{Alice\ sees}\ 0| + |\beta|^{2}|\mathrm{Alice\ sees}\ 1\rangle \langle \mathrm{Alice\ sees}\ 1 |\\
\rho_{B} &=& |\alpha|^{2}|\mathrm{Bob\ sees}\ 0\rangle \langle \mathrm{Bob\ sees}\ 0| + |\beta|^{2}|\mathrm{Bob\ sees}\ 1\rangle \langle \mathrm{Bob\ sees}\ 1 |,
\end{eqnarray}
which returns the standard probabilities.  Considered together, we have:
\begin{eqnarray}
\nonumber \rho_{AB} &=& |\alpha|^{2}|\mathrm{Alice\ sees}\ 0\rangle \langle \mathrm{Alice\ sees}\ 0| |\mathrm{Bob\ sees}\ 0\rangle \langle \mathrm{Bob\ sees}\ 0|\\
&+& |\beta|^{2}|\mathrm{Alice\ sees}\ 1\rangle \langle \mathrm{Alice\ sees}\ 1 | |\mathrm{Bob\ sees}\ 1\rangle \langle \mathrm{Bob\ sees}\ 1|,
\end{eqnarray}
\end{widetext}
which reproduces the standard correlations (which can also be seen via the vanishing conditional entropy:  $S(A|B) = S(AB) - S(B) = 0$ \---- i.e. if we know the results of Bob's experiment, then we know the results of Alice's experiment, and vice versa).  The no-communication theorem still applies as well as ever, as any local unitary operation applied to Alice alone will have no effect on the information available to Bob, due to the partial trace.

Of course, this discussion of the EPR scenario introduces no new experimentally measurable results \---- it is the same familiar story, merely told in a different language.  The novelty is simply that the conceptual baggage created by the concept of instantaneous, non-local collapse can be discarded entirely in favor of local statements about the state of the observers, and shared information that exists on certain constant-time slices.  All operations described by all observers are entirely local from start to finish, and thus Lorentz transformations imply no strange, superluminal propagation of quantum information for observers in differing states of motion.

\section{Unitarity and the time-reversed quantum Zeno effect}

This approach to quantum theory emphasizes the fact that measurement and collapse are merely the information theoretic consequences of a system becoming entangled with an observer, and that all evolution is fundamentally unitary.  A consequence of this fact is that time-reversal becomes a useful symmetry to invoke, even in the presence of quantum measurements.

To see this feature in action, we will analyze a well-known effect of quantum measurement \---- the quantum Zeno effect~\cite{Misra1}.  Ordinarily, the non-unitary formalism of projective measurement is invoked to demonstrate the surprising fact that \emph{merely observing} a quantum system can effectively slow its evolution.  It has also been shown that clever use of projective measurement can also be used to accelerate the evolution of a system, in what has been termed the quantum anti-Zeno effect~\cite{Kofman1} (here, we take the term ``quantum Zeno effect" to be more general than the particular application to decaying atomic states).

However, it is clear that projective measurement corresponds to entanglement with an observer.  Hence, both the quantum Zeno effect and the quantum anti-Zeno effect may be obtained as the result of entangling a quantum system with an ancilla, as we will soon show.  We would further like to point out that the time reversal of these effects also represents a way to manipulate the evolution of a quantum system.  That is, \emph{disentangling} a quantum system from an ancilla can be used to accelerate (or slow) its evolution.  We refer to this as the time-reversed quantum Zeno effect.  Not surprisingly, a combination of Zeno effect and time-reversed Zeno effect cancel completely, leaving ordinary time evolution.

Let us first analyze the quantum Zeno effect within this framework (here for convenience and familiarity we revert to non-covariant language, where evolution in an external time variable is assumed).  We consider a system composed of three subsystems, $Q$, $A$, and $B$.  $Q$ will be the time-dependent system under study, while $A$ (for ``Alice," or ``ancilla") represents the system we will entangle with $Q$.  Then $B$ (Bob), represents the macroscopic observer, modeled quantum mechanically.  To keep our demonstration as simple as possible, all systems will be modeled by a single qubit, and we take $Q$ to be the only system with non-trivial time evolution in the absence of interactions.  

Specifically, we take $Q$ to evolve independently as $|Q(t)\rangle = \cos(\omega t) |0\rangle + i\sin(\omega t) |1\rangle$, as is well-known to occur when the energy eigenstates of $Q$ are $\frac{1}{\sqrt{2}}(|0\rangle + |1\rangle)$ and $\frac{1}{\sqrt{2}}(|0\rangle - |1\rangle)$.  The systems $A$ and $B$ are taken to be in the state $|0\rangle$ at $t=0$.

In the absence of intermediate entanglement, the situation is straightforward.  At time $t=2\epsilon$ (with $\epsilon$ a small time interval compared to $\omega^{-1}$), Bob interacts with $Q$ by means of a CNOT gate (effectively an ideal measurement), producing the entangled state $|QB(2\epsilon)\rangle = \cos(\omega 2\epsilon) |0\rangle|0\rangle + i\sin(\omega 2\epsilon) |1\rangle |1\rangle$.  At this time, the reduced density operator describing Bob is $\rho_{B} = \cos^{2}(\omega 2\epsilon)\: |0\rangle\langle0| + \sin^{2}(\omega 2\epsilon)\: |1\rangle\langle1|$, and so the probability for Bob to detect $Q$ in the state $|1\rangle$ is thus $\sin^{2}(\omega 2\epsilon) \approx 4\omega^{2}\epsilon^{2}$.

Now we analyze the situation again, with the introduction of an intermediate CNOT operation at $t=\epsilon$, entangling $Q$ with $A$ before Bob's measurement at $t = 2 \epsilon$.  Thus at $t=\epsilon$, we have $|QA(\epsilon)\rangle = \cos(\omega \epsilon) |0\rangle|0\rangle + i\sin(\omega \epsilon) |1\rangle |1\rangle$, while Bob remains in the state $|0\rangle$.  At time $t=2\epsilon$, though, (just before we apply the CNOT operation to entangle Bob with $QA$), this state has evolved to:
\begin{widetext}
\begin{eqnarray}
|QA(2\epsilon)\rangle = \cos(\omega \epsilon)\left[\cos(\omega \epsilon) |0\rangle + i\sin(\omega \epsilon)|1\rangle \right] |0\rangle + i\sin(\omega \epsilon) \left[ \cos(\omega \epsilon) |1\rangle - i\sin(\omega \epsilon)|0\rangle \right] |1\rangle.
\end{eqnarray}

Now applying the CNOT between $Q$ and $B$, representing Bob's measurement, and rearranging gives us the state from which we can extract predictions:
\begin{eqnarray}
|QAB(2\epsilon)\rangle = \cos^{2}(\omega \epsilon) |0\rangle|0\rangle|0\rangle + \sin^{2}(\omega \epsilon) |0\rangle|1\rangle|0\rangle + i\cos(\omega \epsilon)\sin(\omega \epsilon)|1\rangle(|0\rangle + |1\rangle)|1\rangle.
\end{eqnarray}
Performing a partial trace over $Q$ and $A$ gives us the reduced density operator describing Bob:
\begin{eqnarray}
\rho_{B} = \left[ \cos^{4}(\omega \epsilon) + \sin^{4}(\omega \epsilon)\right]|0\rangle \langle0| + 2\cos^{2}(\omega \epsilon)\sin^{2}(\omega \epsilon) |1\rangle \langle1|.
\end{eqnarray}

The probability that Bob finds $Q$ to have transitioned to the $|1\rangle$ state is thus $2\cos^{2}(\omega \epsilon)\sin^{2}(\omega\epsilon) \approx 2\omega^{2}\epsilon^{2}$ \---- half of its value in the case of no intermediate entanglement.  This is the quantum Zeno effect on simply evolving qubits (which can, of course, be iterated).  Note that this is \emph{not} a result of non-unitary projections, or of altering the single particle evolution \---- we have merely introduced an entangling CNOT operation, and nothing else.

As a consequence of avoiding the non-unitarity inherent in the projective measurement formalism, we can identify a new effect simply by invoking time reversal symmetry.  That is, we start with an entangled $QA$ state and then \emph{disentangle} $A$ from $Q$ (e.g. by applying another CNOT operation), with the result that $Q$'s effective evolution is accelerated.  To see this explicitly, we start with the following entangled state as a function of time:
\begin{eqnarray}
|QA(t)\rangle = \alpha \left[\cos(\omega t)|0\rangle + i\sin(\omega t) |1\rangle \right]|0\rangle + \beta \left[\cos(\omega t)|1\rangle - i\sin(\omega t)|0\rangle \right] |1\rangle.
\end{eqnarray}
If Bob were to perform measurements on this system, he would find that $Q$ oscillates from states of maximum probability in the $|0\rangle$-state at time $0, \frac{\pi}{\omega}, \frac{2\pi}{\omega},...$ (for $|\alpha|^{2} > |\beta|^{2}$) to states of maximum probability in the $|1\rangle$-state at times $\frac{\pi}{2\omega}, \frac{3\pi}{2\omega},...$.

However, let us apply a CNOT operation between $Q$ and $A$ before Bob makes his measurement.  We apply the CNOT at time $t=0$.  The now-separable state as a function of time can be described by:
\begin{eqnarray}
|QA(t)\rangle = \left\{ \left[ \alpha \cos(\omega t) - \beta i \sin(\omega t)\right] |0\rangle + \left[ \alpha i\sin(\omega t) + \beta \cos(\omega t) \right]|1\rangle \right\} |0\rangle.
\end{eqnarray}
\end{widetext}
After the disentangling CNOT, the solution continues to oscillate with frequency $\omega$ between states of maximum probability in $|0\rangle$ and states of maximum probability in $|1\rangle$, but the peaks and troughs will no longer generally occur at $t= 0, \frac{\pi}{2\omega}, \frac{\pi}{\omega}, \frac{3\pi}{2\omega}, \frac{2\pi}{\omega},...$.  In the case that $\alpha = \cos(\theta)$ and $\beta = -i\sin(\theta)$ for any small value of $\theta$, we have exactly the time-reversal of the Zeno effect shown above, and we find that evolution has been advanced precisely by an amount $\delta t = \frac{\theta}{\omega}$, leaving us with the solution:
\begin{eqnarray}
|Q(t)\rangle = \cos(\omega (t + \delta t))|0\rangle + i\sin(\omega (t + \delta t))|1\rangle 
\end{eqnarray}

Note that this works also for negative values of $\theta$, in which case we have effectively time-reversed the \emph{anti}-Zeno effect (i.e. we have delayed evolution by disentanglement).

Thus we see that through measurement-like interactions (CNOT operations, in our analysis), the quantum Zeno effect essentially represents a trade of evolution for entanglement.  But in the absence of projective collapse, the logic can be reversed and we can also trade entanglement for evolution. 

\section{A note on probabilities in quantum cosmology}
Before we conclude, we would like to emphasize one more feature of covariant quantum information.  That is, it is completely self-contained.  By this, we mean that the theory does not require an external, classical observer in order to properly interpret probabilities.  In fact, the theory requires that we avoid such constructs \---- probabilistic predictions are not made in CQI unless the observer is included as part of the full description of the system.

This seems to suggest a natural application with quantum cosmology, where the meaning of quantum probabilities is controversial for exactly this reason \---- to whom do probabilities apply?  We would like to suggest that some simple cosmological models (e.g. the simplest minisuperspace cosmologies) simply do not have a consistent probabilistic interpretation, because they do not include enough degrees of freedom to model a quantum observer, and do not model an entangling interaction required to define a measurement.  We defer a detailed analysis of this proposal to future work, but we note that it seems to mesh perfectly with historic insights into the character of quantum gravity and quantum cosmology.  Consider, for example, the following quote by DeWitt~\cite{DeWitt1}:
\begin{quotation}
Perhaps the most impressive fact which emerges from a study of the quantum theory of gravity is that it is an extraordinarily economical theory.  It gives one just exactly what is needed in order to analyze a particular physical situation, but not a bit more.  Thus it will say nothing about time unless a clock to measure time is provided, and it will say nothing about geometry unless a device (either a material object, gravitational waves, or some other form of radiation) is introduced to tell when and where the geometry is to be measured.
\end{quotation}

Our proposal merely takes this insight a step further \---- a fully consistent, generally covariant quantum theory says nothing about \emph{probabilities} unless a quantum subsystem (playing the role of the observer) is introduced to become entangled with the degrees of freedom being studied.  Once this is done, however, interpretational problems go away \---- the probabilities are \emph{not} the probability that the entire universe will collapse to some specific state.  The probabilities simply refer to the mixed state of the observer alone, which is in principle the best description of the observer quantum theory can offer.

\section{Conclusions}

The picture painted by covariant quantum information is most remarkable in the way it ties together and clarifies seemingly unrelated issues in a single, coherent framework.  Who would have expected that making a quantum measurement formalism compatible with general covariance would immediately result in a formalism that is also automatically realist, unitary, local, and which defines a quantum mechanical arrow of time independent of any pre-existing causal structure or time coordinate?  Furthermore, the theory suggests a new class of readily measurable quantum mechanical effects which are obtainable simply by invoking time-reversal symmetry in the presence of standard, measurement-related phenomena (the first example of these being the time-reversed quantum Zeno effect).  

We would like to acknowledge useful discussions with Christoph Adami and John Sipe, as well as support from the Intelligence Advanced Research Projects Activity and the Army Research Office.

\bibliography{ref}

\end{document}